\documentclass[aps,prb,twocolumn,groupedaddress]{revtex4}

\usepackage{graphicx}
\usepackage{amssymb}

\begin{document}


\title{Multiple phase slips phenomena in mesoscopic superconducting rings}


\author{Mathieu Lu-Dac and V. V. Kabanov}
\affiliation{Jozef Stefan Institute, Jamova 39, SI 1000 Ljubljana, Slovenia}


\date{\today}

\begin{abstract}
We investigate the behavior of a mesoscopic one-dimensional ring in an external
magnetic field by simulating the time dependent Ginzburg-Landau equations with periodic boundary conditions. We analyze the
stability and the different possible evolutions for the phase slip
phenomena starting from a metastable state. We find a stability
condition relating the winding number of the initial solution and
the number of flux quanta penetrating the ring. The analysis of multiple phase slips solutions is based on analytical results and simulations. The role of the ratio of two characteristic times $u$ is studied for the case of a
multiple phase slips transition. We found out that if $u>>1$, consecutive
multiple phase slips will be more favorable than simultaneous ones. If $u<<1$
the opposite is true and we confirm that $u>>1$ is often a necessary condition to reach the ground state. The influence of the Langevin noise on the kinetics of the phase transition is discussed.
\end{abstract}


\maketitle

\section{Introduction}
Nonequilibrium phenomena in superconductors are a challenging area
for both experimental and theoretical research. They are also
crucial for the development of applications as they are the key to
the appearance of resistive states in superconducting samples and
to the possible use of vortex dynamics. In one-dimension, the resistive
phase slip process foreseen by Little \cite{Little67} has
been described quantitatively by the  theory developed by Langer and Ambegaokar \cite{LA67} and extended by McCumber \cite{McCumber68} and Halperin \cite{MH70} (LAMH). The LAMH theory describes thermally activated phase slips. It
evaluates the resistance of a 1D superconductor when driven out of
thermodynamical equilibrium by a voltage or current source. Since
then, this theory has been accepted in rather good agreement with experiments (for a more complete review of
theoretical and experimental works, see Ref. \onlinecite{Tid90}).

More recently, simulations were carried out on different versions
of the time-dependent Ginzburg-Landau (GL) (TDGL) equations in order to
investigate the dynamics of the process. In particular, the case
of the 1D ring has raised interest since it exhibits multiple
metastable states which can be reached by phase slip processes. In the work of Tarlie and Elder \cite{Tar98}, the ring was submitted to an electromotive force
which constantly accelerates the superconducting electrons.
Another approach was used by Vodolazov and Peeters \cite{VodoPeet02}, who increased the
magnetic field gradually. The latter simulations were confirmed by the experiments made in Refs. \onlinecite{VodoDubo03} and \onlinecite{MichotteVodo04}.

\begin{figure}
                \includegraphics[width=6cm]{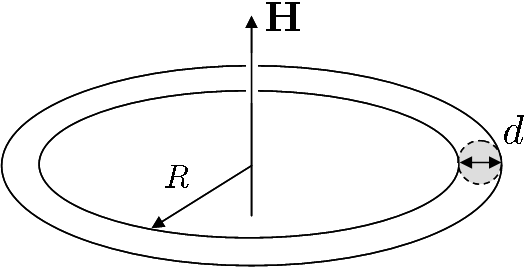}
                \caption{Geometry of the ring of radius $R$ and thickness $d$ with the magnetic field $\mathbf{H}$.}
               \label{fig:ring}
             \end{figure}

In the present work, we consider a superconducting ring of
thickness $d$, radius $R$ and length $L$ as represented in Fig. \ref{fig:ring}. For simplicity, we consider
$d<<\xi<<\lambda_{\text{eff}}$, $R \gtrsim \xi$ and
$R<<\lambda_{\text{eff}}$, where $\xi$ is the coherence length and
$\lambda_{\text{eff}}$ is the Pearl \cite{Pearl64} penetration depth. The
first two conditions allow us to treat the ring as one dimensional
and the last two conditions account for the mesoscopic size of the
ring. We apply an external constant magnetic field $\mathbf{H}$,
perpendicular to the 1D ring.
The field $\mathbf{H}$ is an external parameter that controls the parameters
(superfluid density, current etc.) of the ring.

Our approach is different from that of Refs. \onlinecite{Tar98} and \onlinecite{VodoPeet02} because we investigate the stability of a stationary
state and observe the relaxation process from the initial
metastable state to a stable one, without any evolving external parameter
influencing the dynamics. Indeed, starting from a solution of the
stationary GL equations, we derived the stability condition
regarding small time-dependent perturbations. For example starting from a stable
state, an increase in the magnetic field can result in the
superconducting state to evolve toward a new equilibrium. This new equilibrium is found via one or more phase slip processes. Moreover, using an instant increase in the magnetic field in contrast with the ramp applied in Ref. \onlinecite{VodoPeet02} allows us to investigate the competition between consecutive and simultaneous phase slips.

\section{The time dependent equations}

The TDGL equations have been derived in different ways in order to
describe the different nonequilibrium properties of superconductors. The range of validity of the TDGL equations has been widely discussed in the literature (see, for example, Refs. \onlinecite{GK75} and \onlinecite{IvlevKopnin84}). It requires not only slow variation in the order parameter in space in comparison with the diffusion length but also slow variation in the order parameter in time in comparison with the inelastic scattering time. Therefore the TDGL equations provide an accurate description of kinetics if the size of the sample is larger or comparable with the coherence length $\xi$ and if the time scale is larger than the inelastic-scattering time. Here we use the simplest version that allows to study the phase slip phenomena. It takes into account the presence of magnetic field and
the possibility of charge imbalance by using the vector
potential defined in the symmetric gauge and the electrostatic potential as introduced by Gor'kov and Kopnin.\cite{GK75} The first equation in dimensionless units reads as:
\begin{equation}
 u\left (\frac{\partial \Psi}{\partial t}+i\Phi\Psi\right )
            =\Psi-\Psi|\Psi|^2-(i\nabla+\mathbf{A})^2\Psi + \eta. \label{eq:dtdgl1noise}
\end{equation}
The dimensionless order parameter $\Psi =\rho( \mathbf{r} , t )e^{i\theta( \mathbf{r} ,
    t)}$ takes the value $\Psi=1$ in the equilibrium and in the absence of field. The space and time coordinates $(\mathbf{r},t )$ are measured in units of the coherence length $\xi$ and of the characteristic relaxation time of the phase $\tau_{\theta}=\frac{4\pi\lambda_{\text{eff}}^2\sigma_n}{c^2}$ ($\sigma_n$ is the conductivity of the normal state and $c$ is the speed of light). The Pearl penetration depth $\lambda_{\text{eff}}=\frac{\lambda^2}{d}$
has been used instead of the London penetration depth
$\lambda$ since the thickness $d$ is small. The vector potential $\mathbf{A}$ is written in units of $\frac{\phi_0}{2\pi\xi}$ ($\phi_0$ is the flux quantum) and the electrostatic
    potential
    $\Phi$ in units of $\frac{\hbar}{2e\tau_{\theta}}$, with $e$ being the electron mass and $\hbar$ as the reduced Planck constant.\cite{Schmidt97}

    The only dimensionless parameter left in the equation is the ratio $
u=\frac{\tau_{\rho}}{\tau_{\theta}}$ between the two characteristic times: $\tau_{\rho}$ is the
characteristic time of the evolution of the amplitude of the order
parameter, whereas $\tau_{\theta}$ accounts for the dynamics of
the phase. The estimates from microscopic theories give the value of $u$ ranging from $5.79$ to $12$ (see Refs. \onlinecite{GorElia68} and \onlinecite{Schmidt66}). However, we consider here the general case where $0<u<\infty$.

We also add a Langevin noise $\eta$ during the simulations. The intensity of the noise may vary considerably from one material to another and depends on the experimental conditions. Nevertheless, for the case of the second-order phase transition, the noise should be small and on the order of $\frac{k_B T}{ d^2 H_{c_2}^2 \xi} \sim \frac{T}{T_c} (\frac{k_B T_c}{E_F})^2 <<1$ in dimensionless units when $d \lesssim \xi$. Here, $k_B$ is the Boltzman constant, $T$ is the temperature, $T_c$ is the critical temperature, $H_{c_2}$ is the second critical field and $E_F$ is the Fermi energy.

The first TDGL Eq. (\ref{eq:dtdgl1noise})  is very similar to what is sometimes referred to as the complex GL (CGL) equation, except for the Langevin noise and the terms containing
electrostatic potential and vector potential. The second TDGL equation is the decomposition of the total current into
the superconducting and the normal currents
\begin{eqnarray}
 \nabla\times(\nabla\times  \mathbf{A}) &=&-\frac{1}{\kappa^2} \bigg[\frac{i}{2}(\Psi^*\nabla\Psi-\Psi\nabla\Psi^*)\nonumber \\*
&& +\mathbf{A}|\Psi|^2+\frac{\partial \mathbf{A}}{\partial t}+\nabla\Phi \bigg].  \label{eq:dtdgl2}
\end{eqnarray}
The total current $\mathbf{j}=\nabla\times(\nabla\times
\mathbf{A})$ is the sum of the superconducting current
$\mathbf{j_s}
=-\frac{i}{2}(\Psi^*\nabla\Psi-\Psi\nabla\Psi^*)-\mathbf{A}|\Psi|^2$ and the normal current $\mathbf{j_n}=-\frac{\partial
\mathbf{A}}{\partial t}-\nabla\Phi$ in dimensionless
units. The GL parameter $\kappa=\frac{\lambda_{\text{eff}}} { \xi}$
is large and the dimensions of the ring are small. Therefore, we
neglect the corrections to the vector potential. Moreover, using
the electroneutrality relation $\text{div} \mathbf{j} = 0$ (see Ref. \onlinecite{GK75} for details),
Eq. (\ref{eq:dtdgl2}) is simplified as

\begin{equation}
    \nabla^2\Phi=
            - \nabla\left [\frac{i}{2}(\Psi^*\nabla\Psi-\Psi\nabla\Psi^*)+\mathbf{A}|\Psi|^2 \right ].  \label{eq:dtdgl2bis}
\end{equation}

The stability of the stationary solutions of the CGL equation has been
studied in details (see Ref. \onlinecite{AransKramer02}), and in
1D, one can derive the conditions of Eckhaus-type instability.
In the presence of magnetic field, most of the results are
comparable. Indeed, we can rewrite Eq. (\ref{eq:dtdgl1noise}) in
the same way as in Ref. \onlinecite{KZ85} by separating real and imaginary
parts and neglecting the noise:

\begin{equation}
    u\frac{\partial \rho}{\partial t}=\frac{\partial^2 \rho}{\partial
    x^2}+\rho\left [1-A^2-\rho^2+\frac{\partial \theta}{\partial x}\left (2A-\frac{\partial \theta}{\partial
    x}\right )\right ]\label{eq:realKramer}
\end{equation}
and
\begin{equation}
    u\rho\left (\frac{\partial\theta}{\partial t}+\Phi\right )=\frac{1}{\rho}\frac{\partial }{\partial x}\left [\rho^2\left (\frac{\partial\theta}{\partial
    x}-A\right )\right ],\label{eq:imKramer}
\end{equation}
where $\rho$ is the amplitude, $\theta$ is the phase of the complex
order parameter $\Psi$, and $x$ is the longitudinal dimensionless
spatial coordinate. The superconducting current
$\rho^2(\frac{\partial \theta}{\partial x}-A)$ needs to be uniform
for stationary solutions. It corresponds to zero electrostatic
potential and leads to $\frac{\partial^2 \theta}{\partial x^2}=0$.

We use a similar derivation of the constant solutions as in
Refs. \onlinecite{IvlevKopnin84} and \onlinecite{KZ85} by describing Eqs. (\ref{eq:realKramer}) and
(\ref{eq:imKramer}) as the equations of motion of a particle in a
potential. We obtain a family of solutions in the well-known
twisted plane-wave form,
\begin{equation}
\Psi_k=\sqrt{1-(A-k)^2}e^{i(kx+\theta_0)},\label{eq:soldimless}
\end{equation}
where $k$ and $\theta_0$ are real numbers. Periodic boundary
conditions imply that $k=\frac{2n\pi}{l}$, where $l$ is
the dimensionless length of the wire and $n$ is an integer that
corresponds to the winding number or vorticity of the solution.
Indeed, we have $n=\frac{1}{2\pi}\oint\frac{d\theta}{d x}$. These
solutions are only possible when $j_s< j_c=2/\sqrt{27}$, $j_c$
being the GL critical current.

The stability of such solutions has already been studied in
different cases, mostly in the context of the CGL equation (without including electromagnetic fields). In our case, we
include in the analysis both the vector potential and
the electrostatic potential. We then analyze the stability of a
solution in the form of Eq. (\ref{eq:soldimless}) by linearizing the
equations disturbed by a perturbation $y(x,t)=y_r(x,t)+iy_i(x,t)$
and performing Fourier transform $\hat{y_r}(q,t) +
i\hat{y_i}(q,t)$. We obtain the system of algebraic equations,

\begin{widetext}

\begin{equation}
            \begin{array}{cccc}
                \left(
                \begin{array}{c}
                    \dot{\hat{y_r}}\\
                    \dot{\hat{y_i}}
                 \end{array}
                \right)
                &=\frac{1}{u}&
                \left(
                \begin{array}{cc}
                    -2[1-(k-A)^2)]-q^2&2i(k-A)q\\
                    -2i(k-A)q-\frac{2i(k-A)u}{q}[1-(k-A)^2]&-q^2-u[1-(k-A)^2]
                \end{array}
                \right)
                \left(
                \begin{array}{c}
                    \hat{y_r}\\
                    \hat{y_i}
                \end{array}
                \right).
            \end{array}
\end{equation}
The eigenvalues are consistent with the ones found with neither electrostatic potential nor magnetic field \cite{Tar98} and in the two-dimensional case.\cite{SoininenKopnin94} Only one of the two families of eigenvalues of this system can be positive,
\begin{eqnarray}
 \lambda&=&\frac{1}{2}\sqrt{4-4u+u^2+(-8+16q^2+24u-2u^2)(k-A)^2+(4-20u+u^2)(k-A)^4}\nonumber\\*
           &&-1-q^2-\frac{1}{2}u+\frac{1}{2}u(k-A)^2+(k-A)^2.
\end{eqnarray}
\end{widetext}
Therefore, the stability condition reads as
\begin{equation}
    |k-A|\leq\frac{1}{\sqrt{3}}\sqrt{1+\frac{q^2}{2}}.
\end{equation}
For the case of periodic boundary conditions such as in the
1D ring, we use the Fourier expansion of the
perturbation,
\begin{equation}
\hat{y}(q)=\sum_{m \in \mathbb{Z}} c_m(y)\delta \left (q-\frac{m}{l}\right ),
\end{equation}
where $c_m(y), m\in \mathbb{Z}$ are the Fourier
coefficients of $y$ and $\delta$ is the delta function. Therefore, the stability condition

\begin{equation}
    |k-A|\leq\frac{1}{\sqrt{3}}\sqrt{1+\frac{m^2}{2l^2}} \label{eq:stablefourier}
\end{equation}
is in agreement with those found in previous works.\cite{Tar98, VodoPeet02} Let us rewrite this condition
 in the form,

 \begin{equation}
 m^2\geq l^2\left [6(k-A)^2-2 \right ]. \label{eq:fourierstable}
 \end{equation}
It means that the higher modes corresponding to
multiple simultaneous phase slips will be stable if the length of
the wire is small enough. In other words, the length of the wire can
restrict the number of phase slips that can happen at the same
time.
Moreover, the size of the ring plays an important role in the selection between multiple and consecutive processes as we discuss in Sec. \ref{subsec:multiplePS}. Depending on the size of the ring, different modes can have highest eigenvalue.

\begin{figure}
                \includegraphics[width=7.5cm]{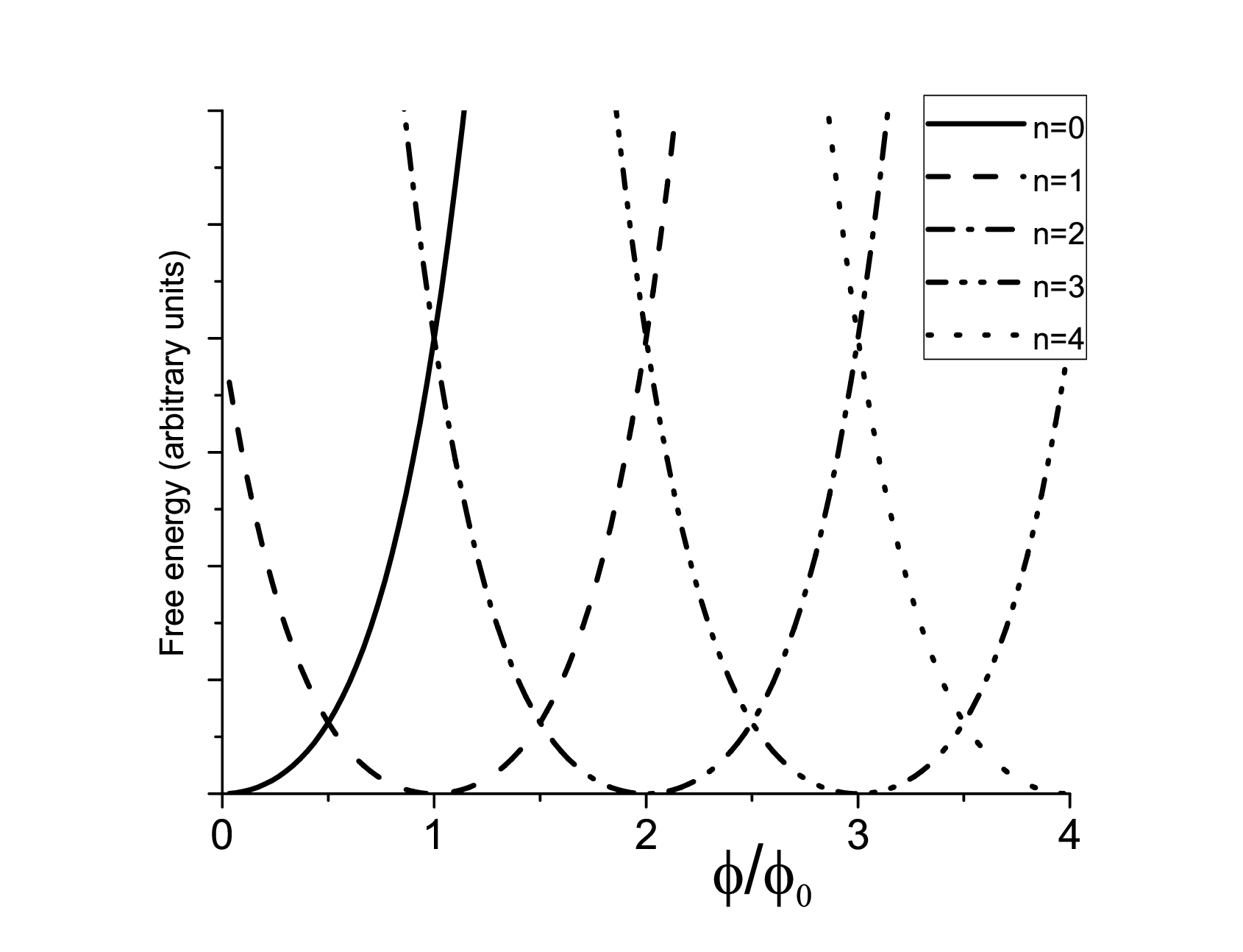}
                \caption{Shape of the free energy of the ring depending on the ratio $\frac{\phi}{\phi_0}$ and the winding number $n$.}
                \label{fig:Freeener}
            \end{figure}

We simplify the
stability condition in:
\begin{equation}
    |k-A|\leq\frac{1}{\sqrt{3}}. \label{eq:stabledimless}
\end{equation}
This condition may be viewed as a generalization of Eq.
(\ref{eq:stablefourier}). It corresponds to the thermodynamically stable supercurrent
$j_s=\rho^2\sqrt{1-\rho^2}$ (see, for example, Fig. 18. in Ref.
\onlinecite{IvlevKopnin84}). Rewriting the condition (\ref{eq:stabledimless}) in
dimensional units, we find a stability condition relating the
winding number $n$ of the solution with the number $\frac{\phi}{\phi_0}$ of flux quanta
penetrating the ring:

\begin{equation}
\left |n-\frac{\phi}{\phi_0}\right |\leq\frac{R}{\xi\sqrt{3}}.
\label{eq:stabledim}
\end{equation}
This condition is consistent with the ground state found by minimizing the
free energy with a solution in the form of Eq. (\ref{eq:soldimless}). As
seen in Fig. \ref{fig:Freeener}, the ground state is reached when the value of $\left |n-\frac{\phi}{\phi_0}\right |$ is minimal.

\section{ Phase slip Simulations }
\subsection{Mathematical formulation of the problem}\label{subsec:mathform}

According to Eq. (\ref{eq:stabledimless}), the winding
number needs to be changed in order to reach a stable state. The
new solution will be closer to the ground state. Therefore, the
superconductor should reach a state with lower current and
energy. Indeed, remembering that the superconducting current is
$\mathbf{j_s}=-\mathbf{A}|\Psi|^2-\frac{i}{2}(\Psi^*\nabla\Psi-\Psi\nabla\Psi^*)$,
the transition to a state of lower current  cannot occur without
modifying the phase and hence the value of $k$. We observe a
transition from the solution $\Psi_{k_0}$ to
        a different solution $\Psi_{k_1}$. Because of periodic boundary conditions, the transition from $k_0$ to $k_1$ requires discontinuity in the
phase. Since the
        order parameter is a single-valued function, this
        discontinuity or jump should be of $2\pi n$, $n$ being an integer number.
         Moreover, $\Psi=0$ at the center of the phase slip to fulfill the continuity condition for the order parameter. Therefore, the transition between two
        solutions is a resistive phase slip. Once the slip is achieved, the phase recovers continuously.

        In our simulation, we start with $k_0=0$. We expect the
        evolution to a state
        $k_1=\frac{2\pi n_1}{l}$ after a number $n_1$ of $2\pi$ phase slips (or an equivalent number of greater phase slips)
        corresponding to the final phase:

        \begin{equation}
            \theta=\frac{2\pi n_1 }{l} x +\theta_0. \label{eq:newphase}
        \end{equation}
 If the number
$n_1$ of phase slips is equal to the integer part of the ratio
$\frac{\phi}{\phi_0}$, the final state reached is the
ground state, but as we show, this is not always the case
and it is difficult to predict (see Sec. \ref{subsec:multiplePS} for details).
        In other words, we observe the transition after an infinitesimal perturbation from the solution
        \begin{equation}
            \Psi_0=\sqrt{1-A^2}\label{eq:stationaryk0}
        \end{equation}
        to another solution after $n_1$ phase slips

        \begin{equation}
            \Psi_{\frac{2\pi n_1 }{l}}=\sqrt{1-\left (A-2\pi n_1/l\right)^2}e^{i\left (2\pi n_1x/l+\theta_0 \right )}. \label{eq:stationaryk1}
        \end{equation}
This solution may correspond to the ground state or a new metastable
state of lower energy.

\subsection{The single phase slip}

In the simulations, we use fast Fourier transforms for spatial
derivatives and Runge-Kutta method for time evolution. As
discussed in Sec. \ref{subsec:mathform} and as described in the
LAMH theory, the amplitude of the order parameter vanishes in a narrow region near the phase slip center for a short time (see Fig. 19 in Ref. \onlinecite{IvlevKopnin84}). Then the amplitude relaxes to a
uniform state with a higher value than it had initially.
The phase of the order parameter develops a sharper sinusoidal form,
until the minimum and the maximum disconnect, at the same moment when the amplitude vanishes and
in the same region, as discussed in Ref. \onlinecite{Ludac08}. Afterward, it relaxes to a sawtooth pattern
corresponding to the new state (\ref{eq:newphase}) so that both
amplitude and phase are consistent with the solution
(\ref{eq:stationaryk1}).

The superconducting current behaves similarly to the
amplitude of the order parameter. It vanishes at the same
point and at the same time as the phase slips.
        This region becomes resistive at that moment. Contrary to the amplitude, the process occurs with a decrease in the
        supercurrent, as explained in Sec. \ref{subsec:mathform}.

        According to the Josephson equation $\frac{d\Delta\theta}{dt}=\Delta\Phi$,
                the electrostatic scalar potential and the phase are directly
                connected ($\Delta \theta$ and $\Delta \Phi$ are respectively, the phase and electrostatic potential difference taken between two arbitrary points). As the
        amplitude of the order parameter goes to zero, the average speed of the
        electrons is reduced and thus some of them gather
        before the region of the phase slip. Therefore, there is a deficit of electrons at the
        end of the region and the electric field is created. This
        voltage influences the phase of the order
        parameter, until it makes a slip of $2\pi$ when the amplitude vanishes. Then, the amplitude
        recovers and the voltage relaxes back to zero, as the
        electrons spread again uniformly along the wire. The ``discontinuity'' in the phase remains  (see Fig. 3 in Ref. \onlinecite{Ludac08} ).

         The characteristic values for the process are deceiving because they depend strongly on the
                material and the thickness of the ring.
                For example in NbN we estimate $\tau_{\theta}\approx 1$ps. The total time of the transition between two
                states separated by a single phase is thus of the order of $100$ps and the characteristic voltage appearing in the ring is of
                about $10\mu$V.

\subsection{Multiple phase slips solutions}\label{subsec:multiplePS}

Here we investigate the case of a transition where the stable state
and the initial state are separated by more than one phase slip. The stability condition (\ref{eq:stabledimless}) is insufficient to predict the final state
and the number of phase slips that occur. The best prediction of the number of phase slips is given
 by finding the winding number of the
ground state (see Fig. \ref{fig:Freeener}). In general, the maximum number of phase slips that can happen is equal to $\left |n-\frac{\phi}{\phi_0}\right |$. Starting with $n=0$, a high magnetic flux will thus be a good condition to observe multiple phase slips.
However, we found out that the ground state is reached
only for certain values of the parameter $u$.

\begin{figure*}
           {\includegraphics[width=7cm]{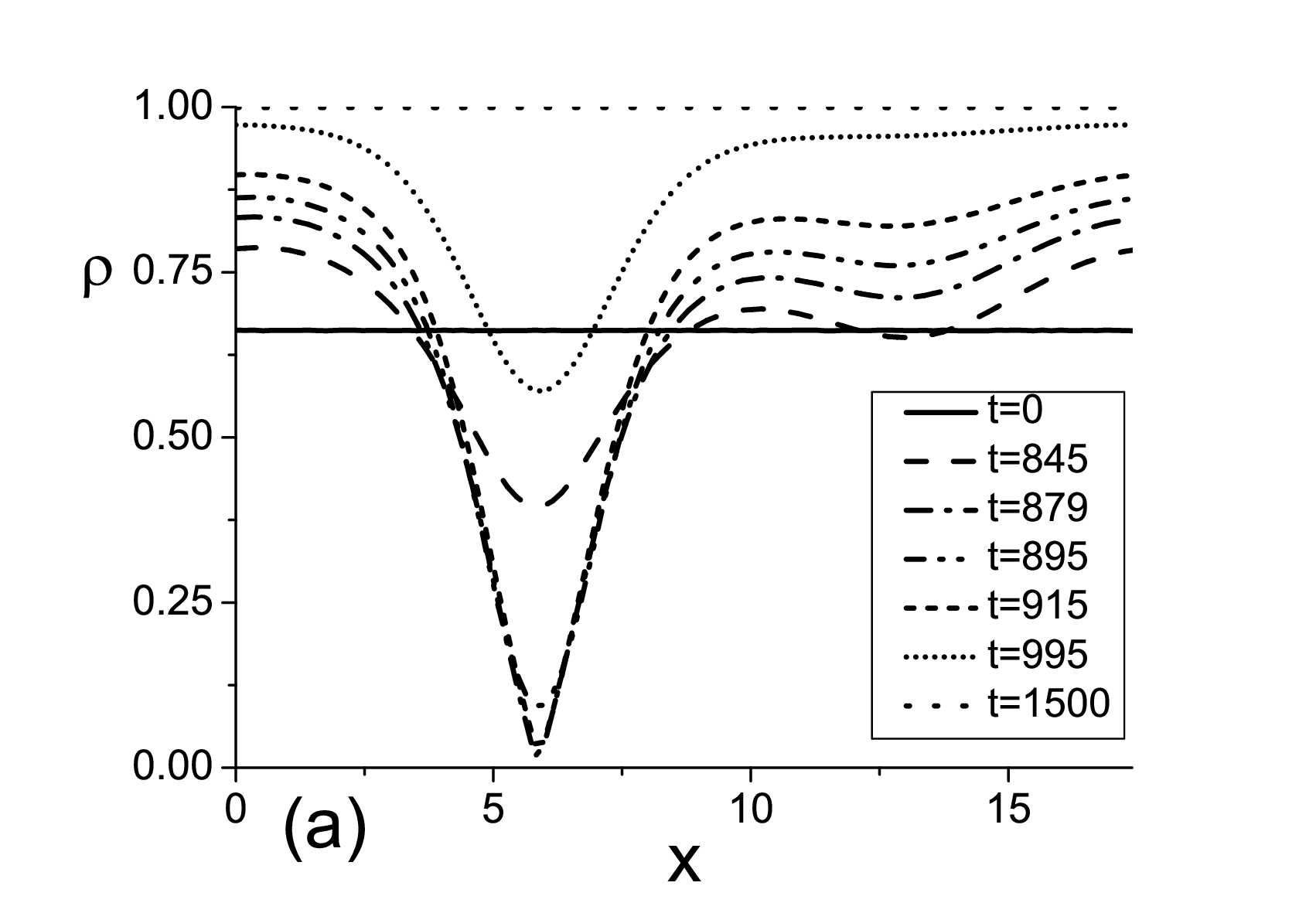}}
            {\includegraphics[width=7cm]{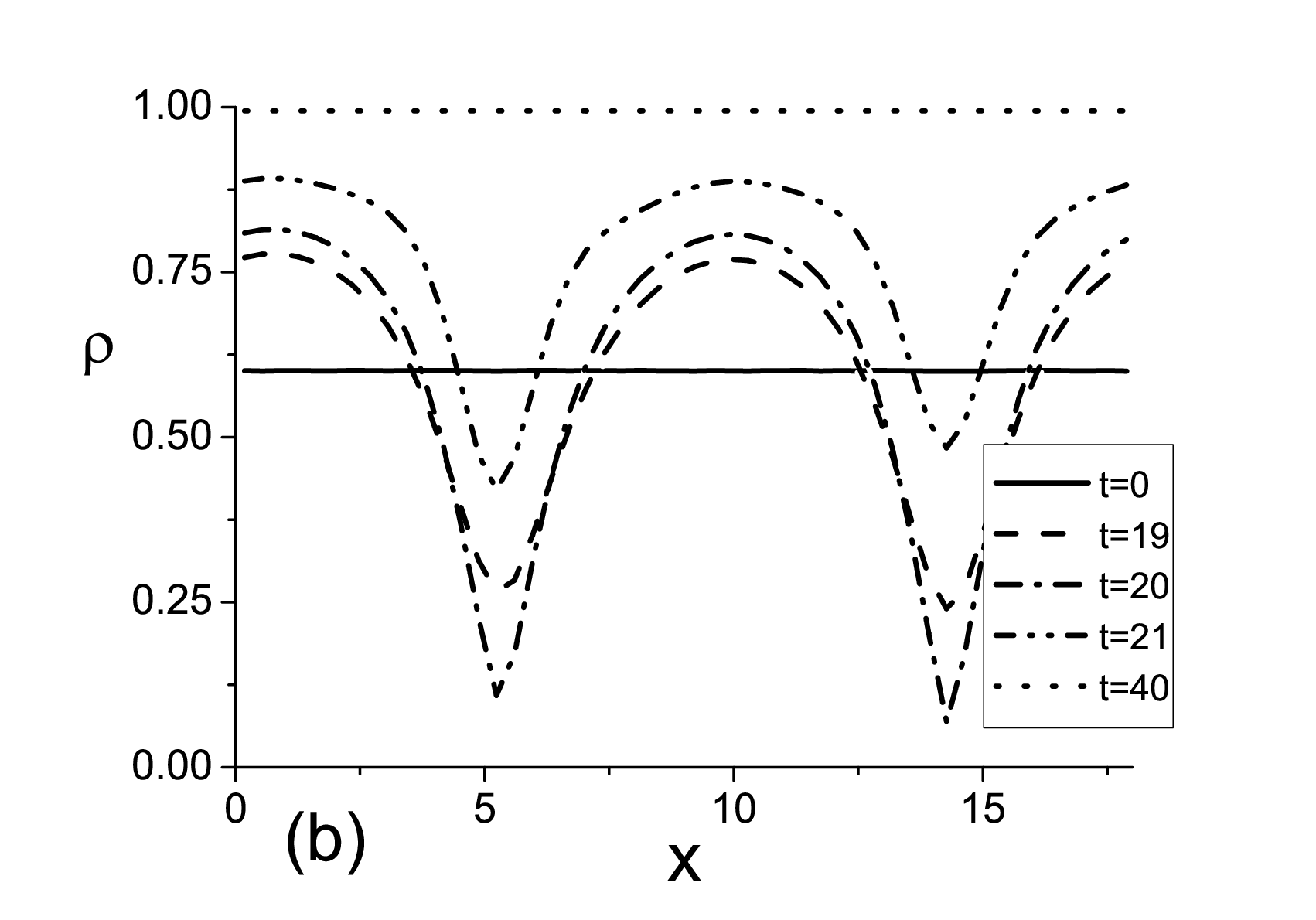}}
            \caption{Evolution of the distribution of the order parameter amplitude
            $\rho$ for two consecutive phase slips (a) and two simultaneous phase slips (b).}
            \label{fig:multiple}
\end{figure*}

If $u>>1$, the phase $\theta$ relaxes faster than the amplitude
$\rho$. It favors multiple phase slips to happen at the same spot, one after another,
as seen in Fig. \ref{fig:multiple}(a). The oscillations of the amplitude of the order parameter during the time of the transition are similar to that observed in Ref. \onlinecite{VodoPeet02}. In this case, the order parameter almost
always relaxes to the ground state, even if an
intermediate state is stable according to condition (\ref{eq:stabledimless}).

If $u<<1$, the amplitude $\rho$ relaxes faster than the
phase. It favors processes where multiple phase
slips happen more or less simultaneously at different phase slip centers as seen in
Fig. \ref{fig:multiple}(b). Therefore, after a certain number of
simultaneous phase slips, the amplitude relaxes to a new uniform
state which will be the final stable state. In that case, the
final state is not necessarily the ground state. These
observations are consistent with the observations made in
Ref. \onlinecite{VodoPeet02} that for $u<<1$ the final state may be
``further away'' from the ground state than for $u>>1$.

\begin{figure*}
           {\includegraphics[width=7cm]{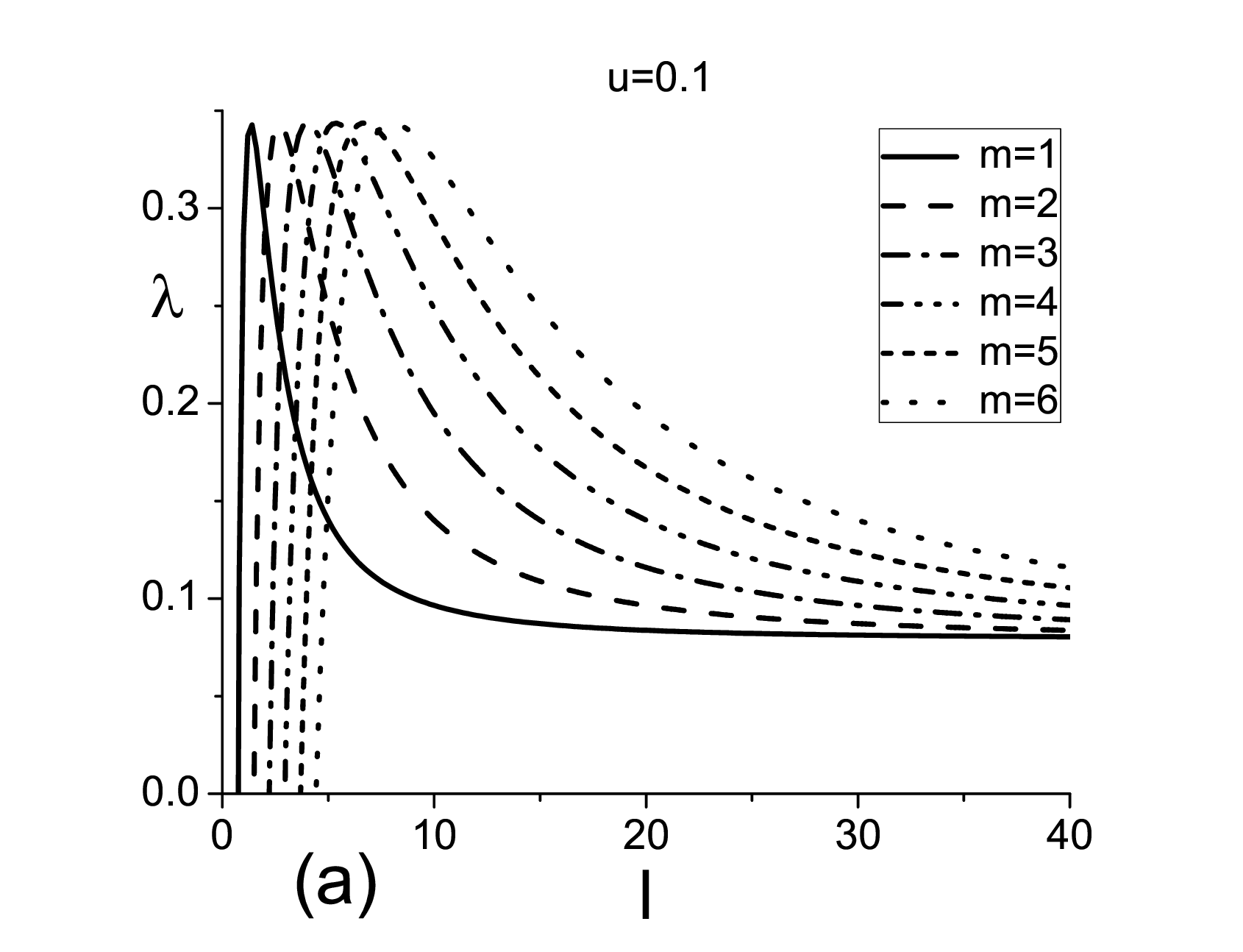}}
            {\includegraphics[width=7cm]{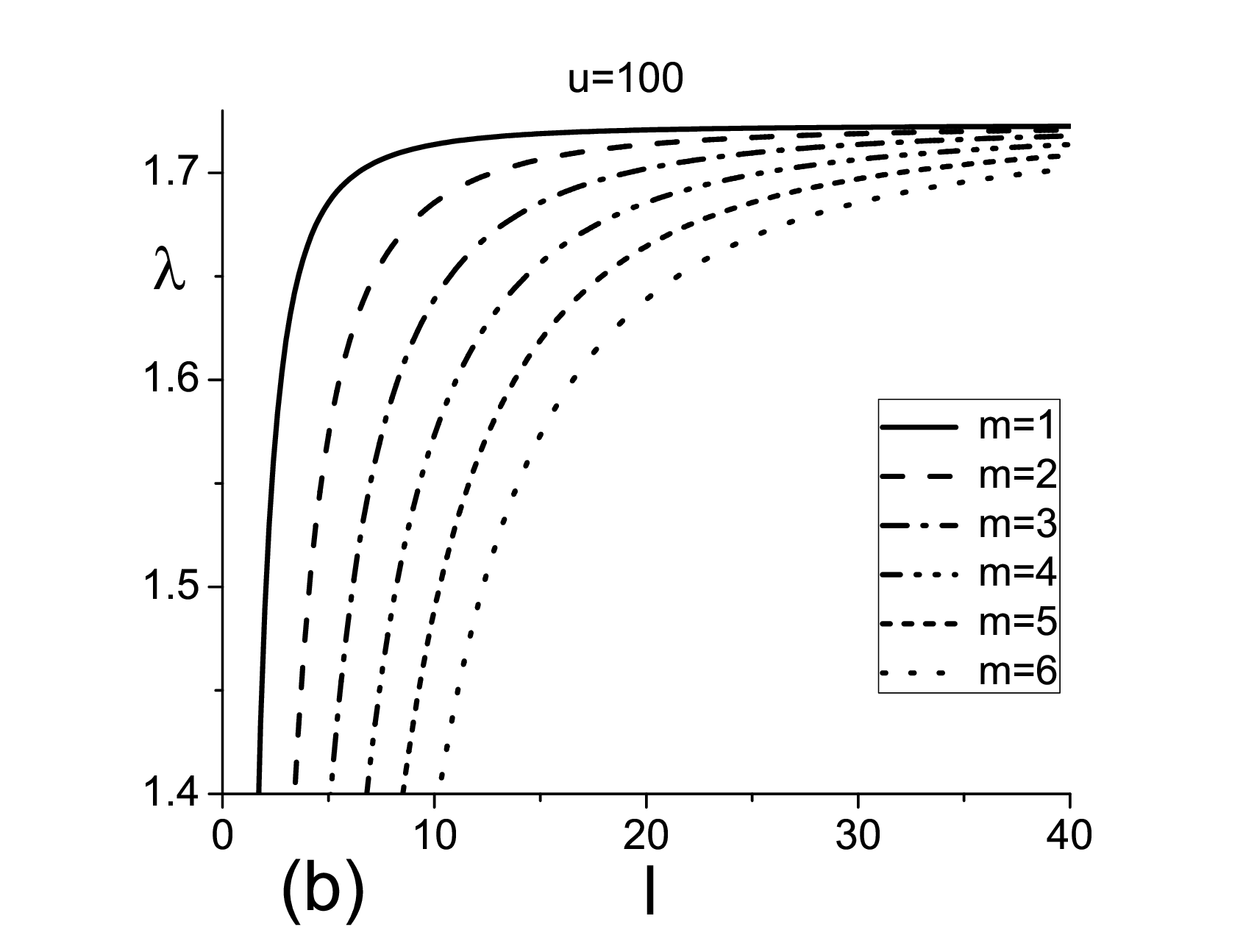}}
           \caption{Eigenvalues $\lambda$ corresponding to different modes $m$ as a function of the size $l$ of the ring for $u=0.1$ (a) and $u=100$ (b). Here $k=0$ and $A=0.8$. When $u>>1$ there is no competition between the modes and consecutive phase slip processes will always dominate. With $u<<1$, the highest eigenvalue depends on the size of the ring. For larger length the simultaneous phase slips will dominate.}
            \label{fig:lambdau}
 \end{figure*}

Mathematically, the importance of the parameter $u$ is emphasized by plotting different eigenvalues as a function of the size of the ring. For $u>>1$, the highest eigenvalues always corresponds to single phase slips happening one at a time. \emph{However, for $u<<1$, different eigenmodes have highest eigenvalue depending on the size of the ring. For large sizes, the simultaneous multiple phase slips modes are the most unstable (see Fig. \ref{fig:lambdau}).}

It is possible to differentiate three cases when multiple phase slips happen: (1) there is only one phase slip center and the amplitude of the order parameter reaches zero several times at the same location before relaxing to a stable state. This is the general consecutive phase slips case. (2) There are more phase slip centers and the amplitude reaches zero at the same time at all locations. (3) There are more phase slip centers but the amplitude reaches zero at a different time for each location.

\begin{figure*}
           {\includegraphics[width=7cm]{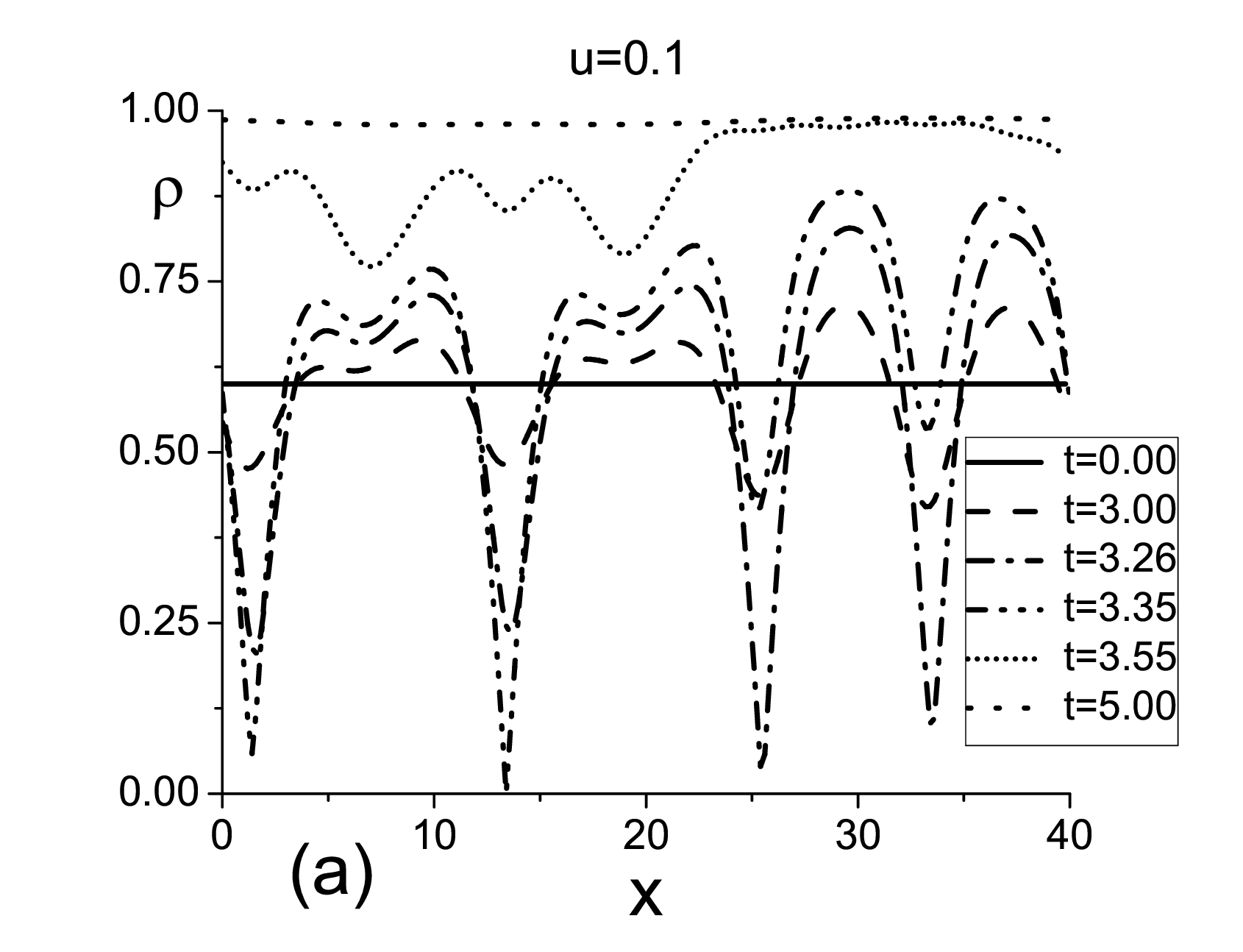}}
            {\includegraphics[width=7cm]{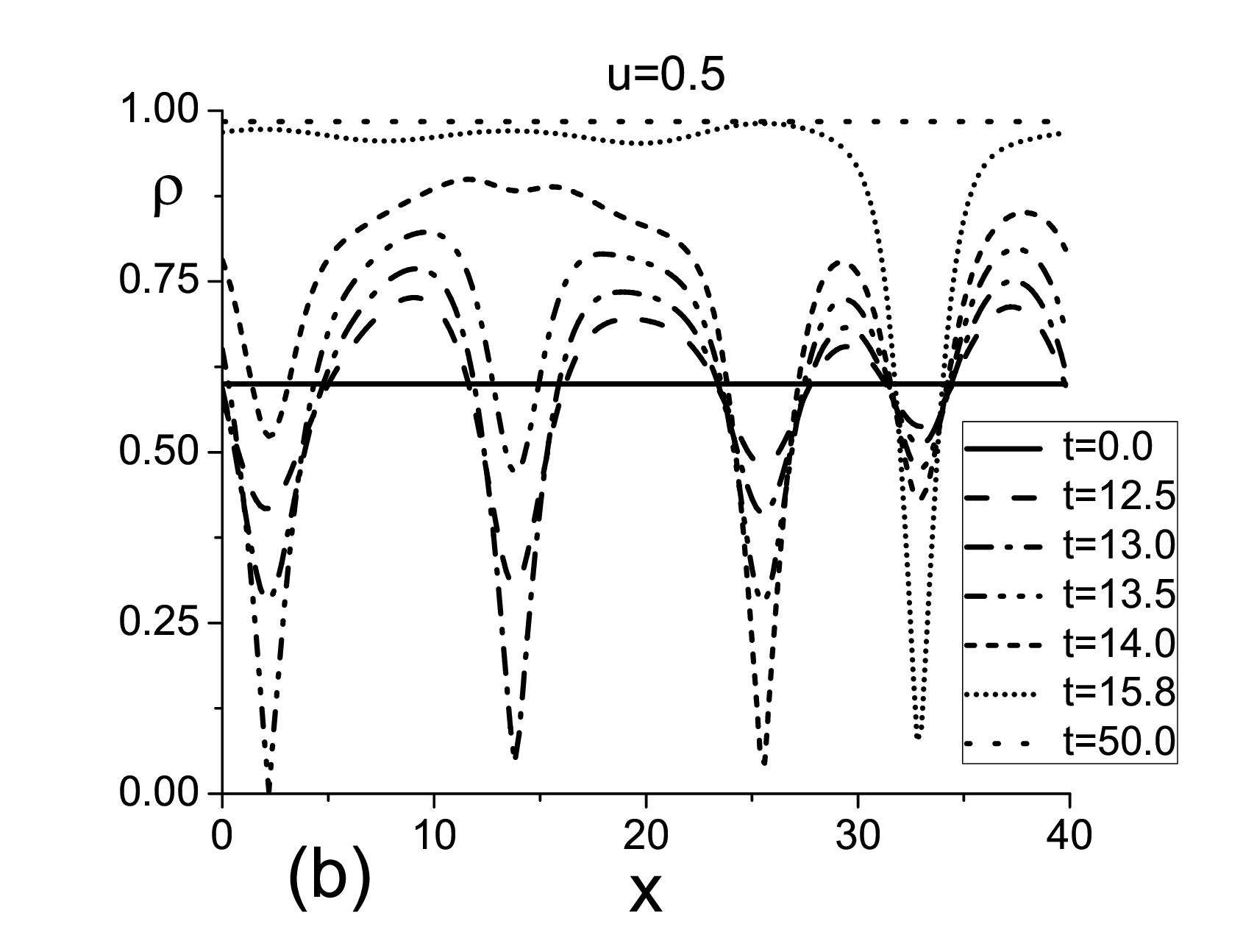}}
            \\
            {\includegraphics[width=7cm]{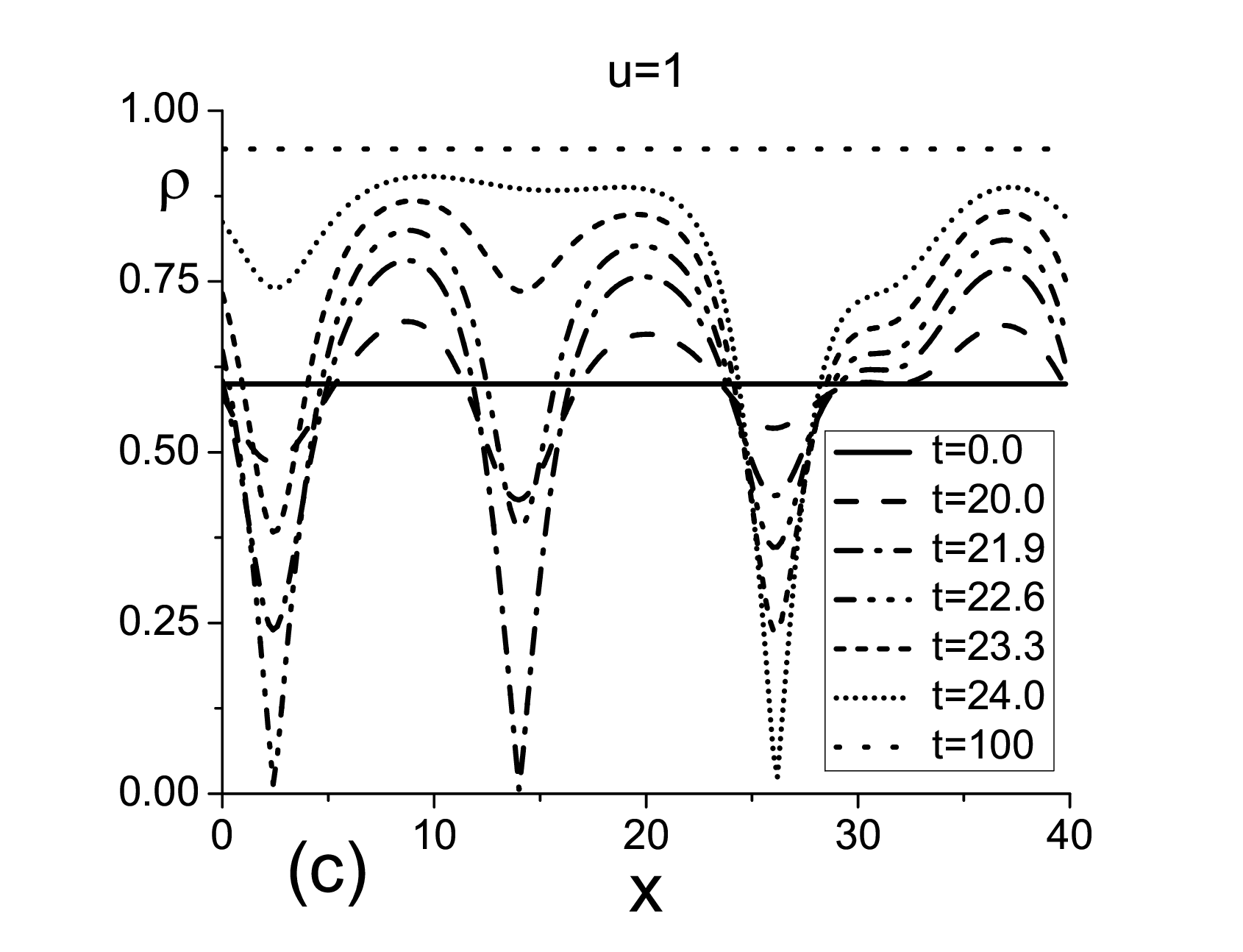}}
            {\includegraphics[width=7cm]{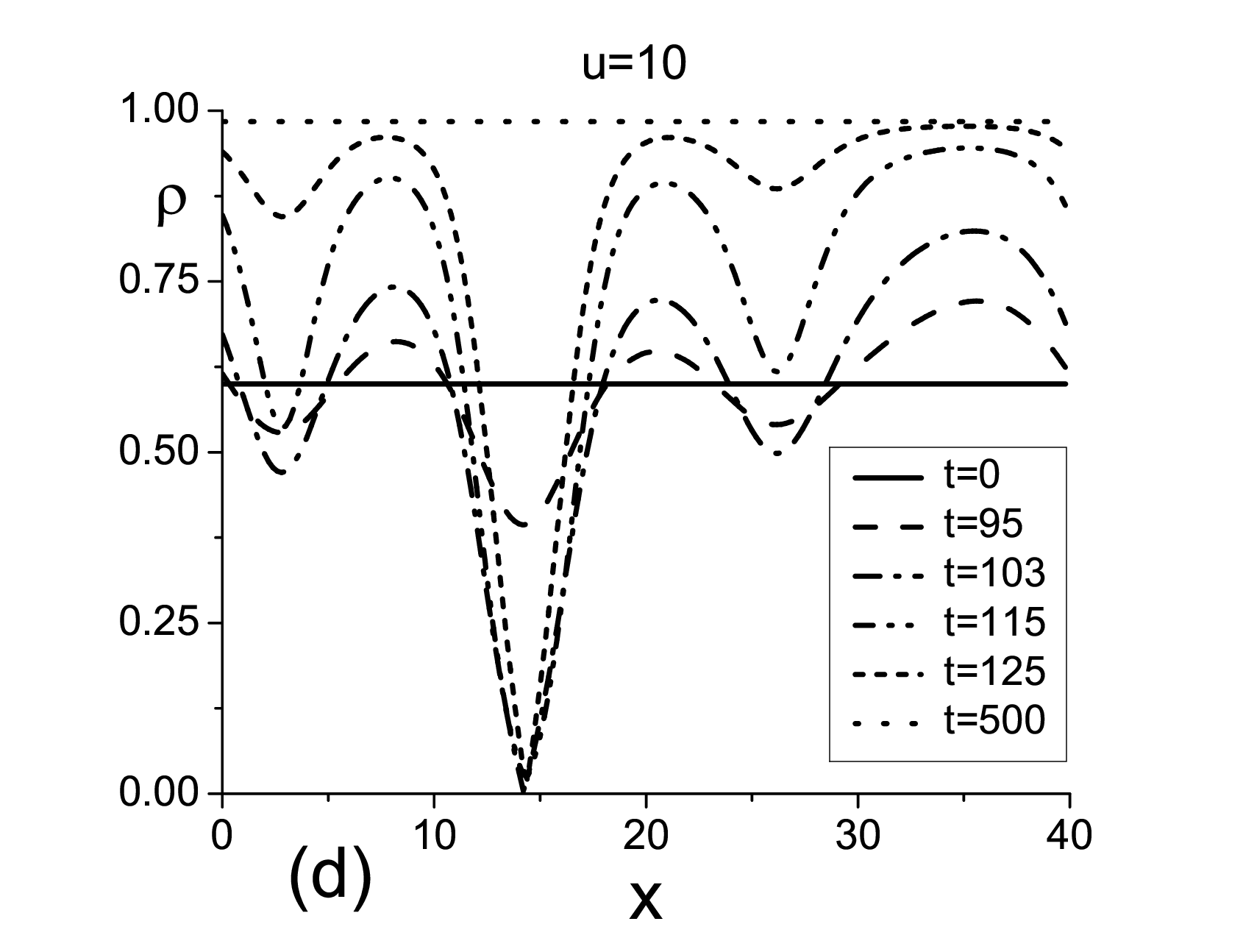}}
            \\
            {\includegraphics[width=7cm]{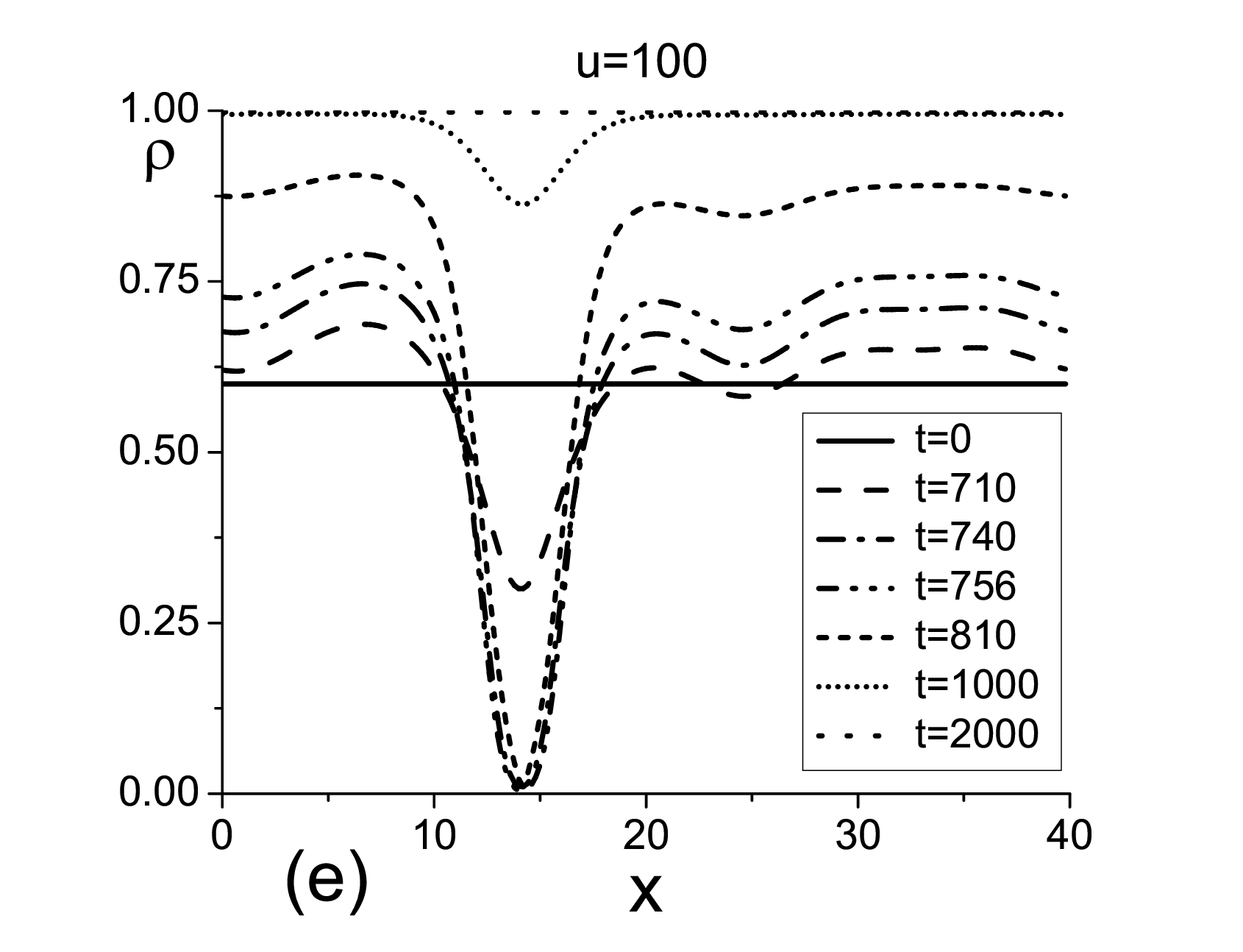}}
            {\includegraphics[width=7cm]{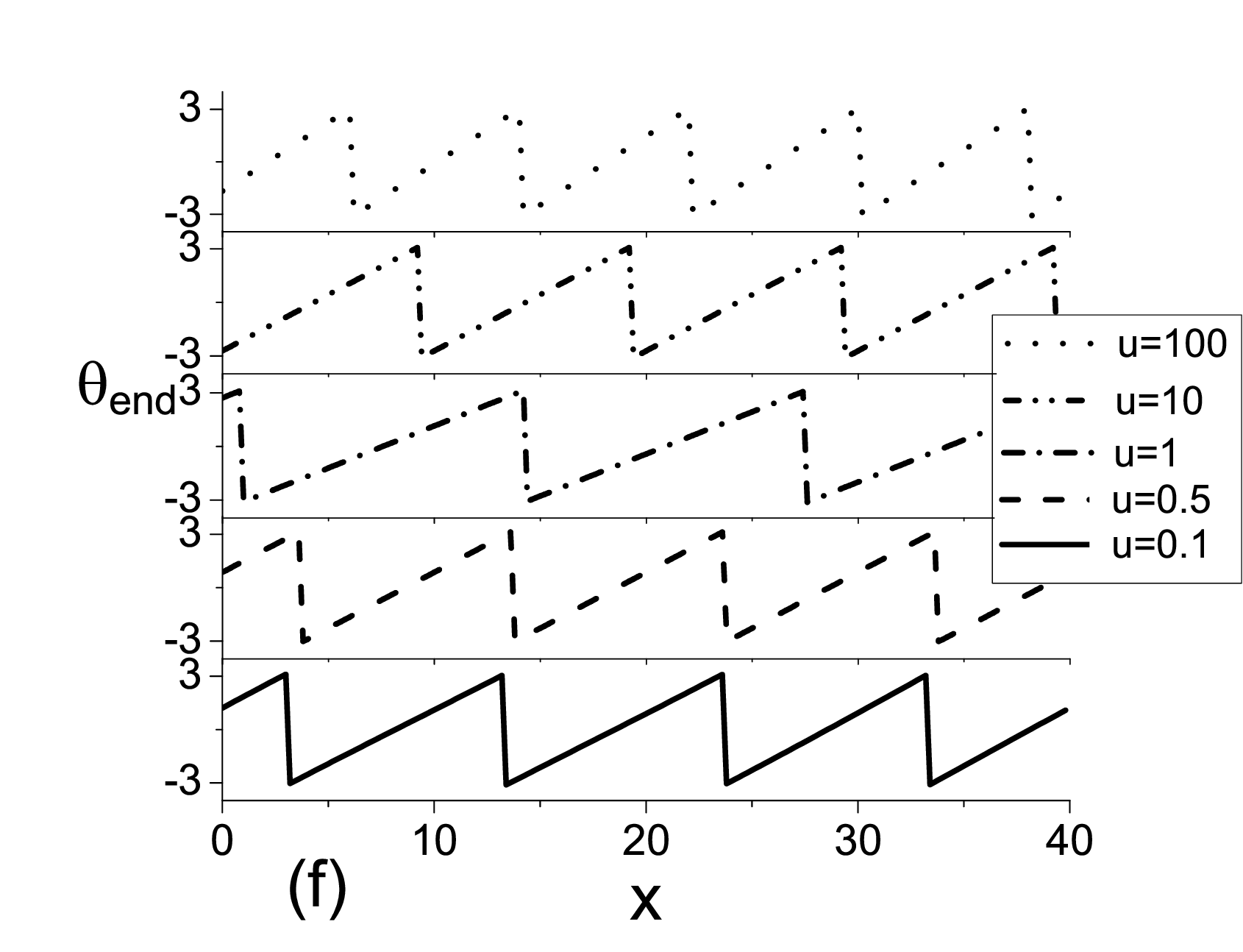}}
            \caption{Evolution of the distribution of the order parameter amplitude $\rho$ (a)-(e)
            exhibiting the competition between simultaneous and consecutive phase slips for different values of $u$ and distribution of the corresponding phase $\theta_{end}$ at the final stable state (f)}
            \label{fig:multipleu}
\end{figure*}

        Here we consider cases (2) and (3) as simultaneous phase slips. Indeed, the amplitude starts to decrease at the same time for all phase slip centers, so that the processes are called simultaneous. In Fig. \ref{fig:multiple}(a) one can also see the competition between simultaneous and consecutive phase slips. Indeed, in that simulation, the amplitude decreases first on two separate spots, but in the end, both phase slips happen consecutively on the same spot. Figure \ref{fig:multipleu} shows the different behaviors depending on the parameter $u$. We confirm that the ground state is only reached for $u>>1$. Indeed, for those simulations, we have $\frac{\phi}{\phi_0}\simeq5$ and we observe five phase slips only when $u=100$. However, the number of phase slips does not necessarily grow monotonically with the value of $u$. Indeed, for $u=1$, we observe less phase slips than for $u=0.5$ and $u=0.1$.

According to Eq. (\ref{eq:fourierstable}) and Fig. \ref{fig:lambdau}, the size of the ring restricts the number of phase slip centers. However, at small sizes, multiple phase slips are often ruled out because the magnetic flux is too low or too high. In particular, the magnetic flux induces more phase slips than the number of phase slip centers allowed by the eigenmodes if
    \begin{equation}
   \left |n-\frac{l A}{2 \pi}\right|>\sqrt{l^2\left [6\left (\frac{2n\pi}{l}-A \right)^2-2\right ]}.
    \end{equation}
   This condition restricts greatly the choice of parameters. In our case, with $k=0$, we need $A<\sqrt{\frac{8\pi^2}{24\pi^2-1}}\simeq 0.578 \approx \frac{1}{\sqrt{3}}  $. Therefore, it limits the simulations to an area which is almost stable and therefore intrinsically not prone to simultaneous phase slips.

\begin{figure*}

           {\includegraphics[width=7cm]{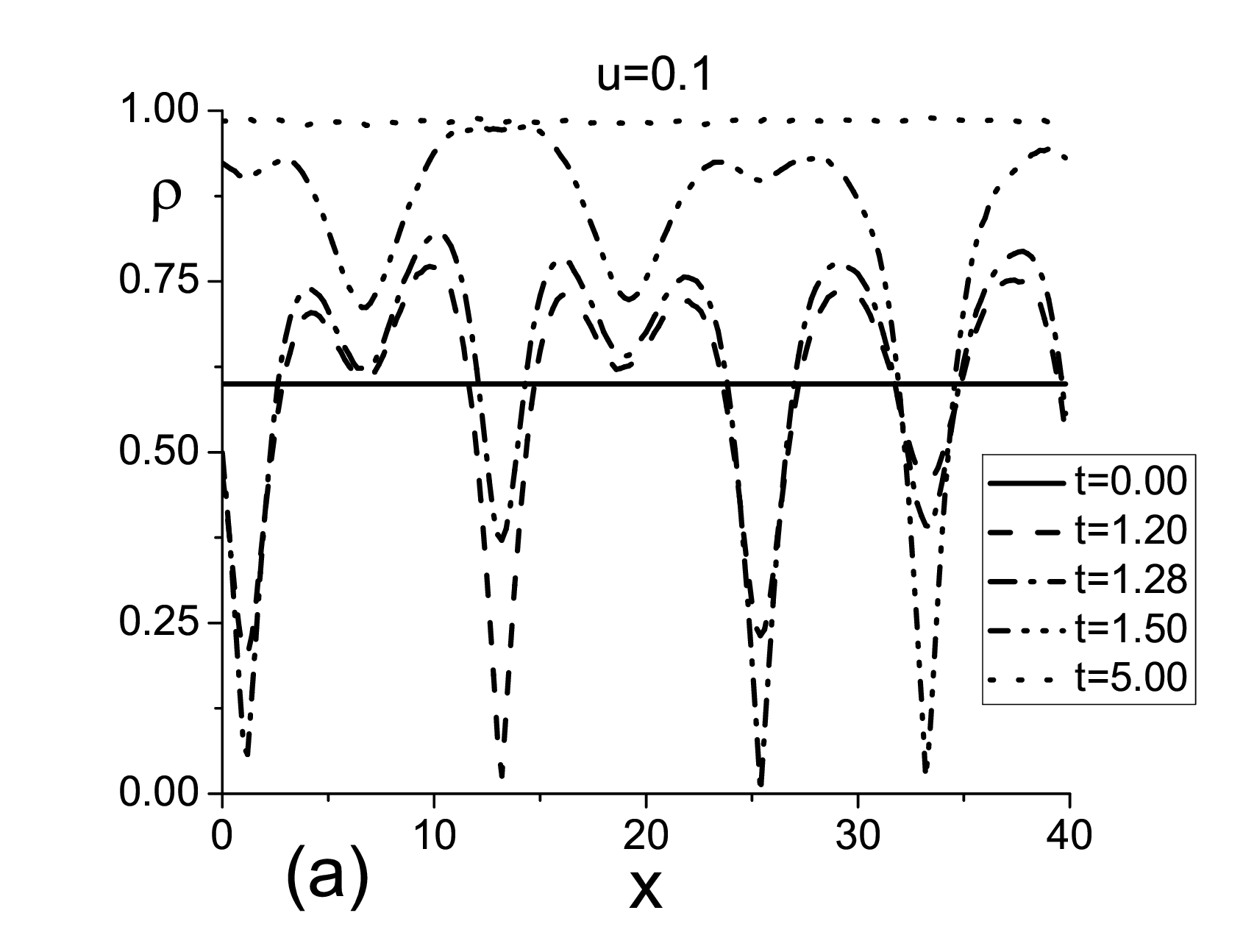}}
            {\includegraphics[width=7cm]{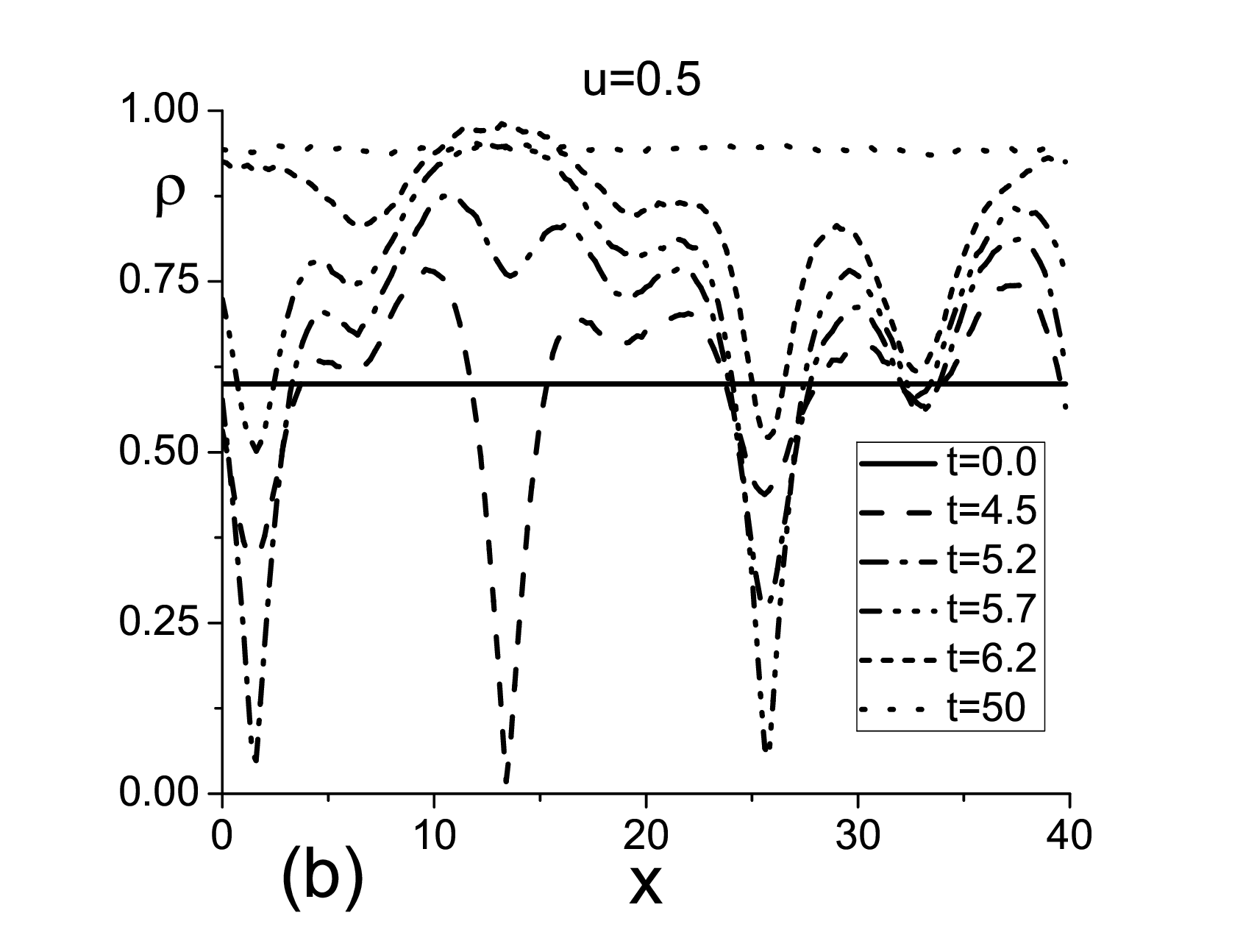}}
            \\
            {\includegraphics[width=7cm]{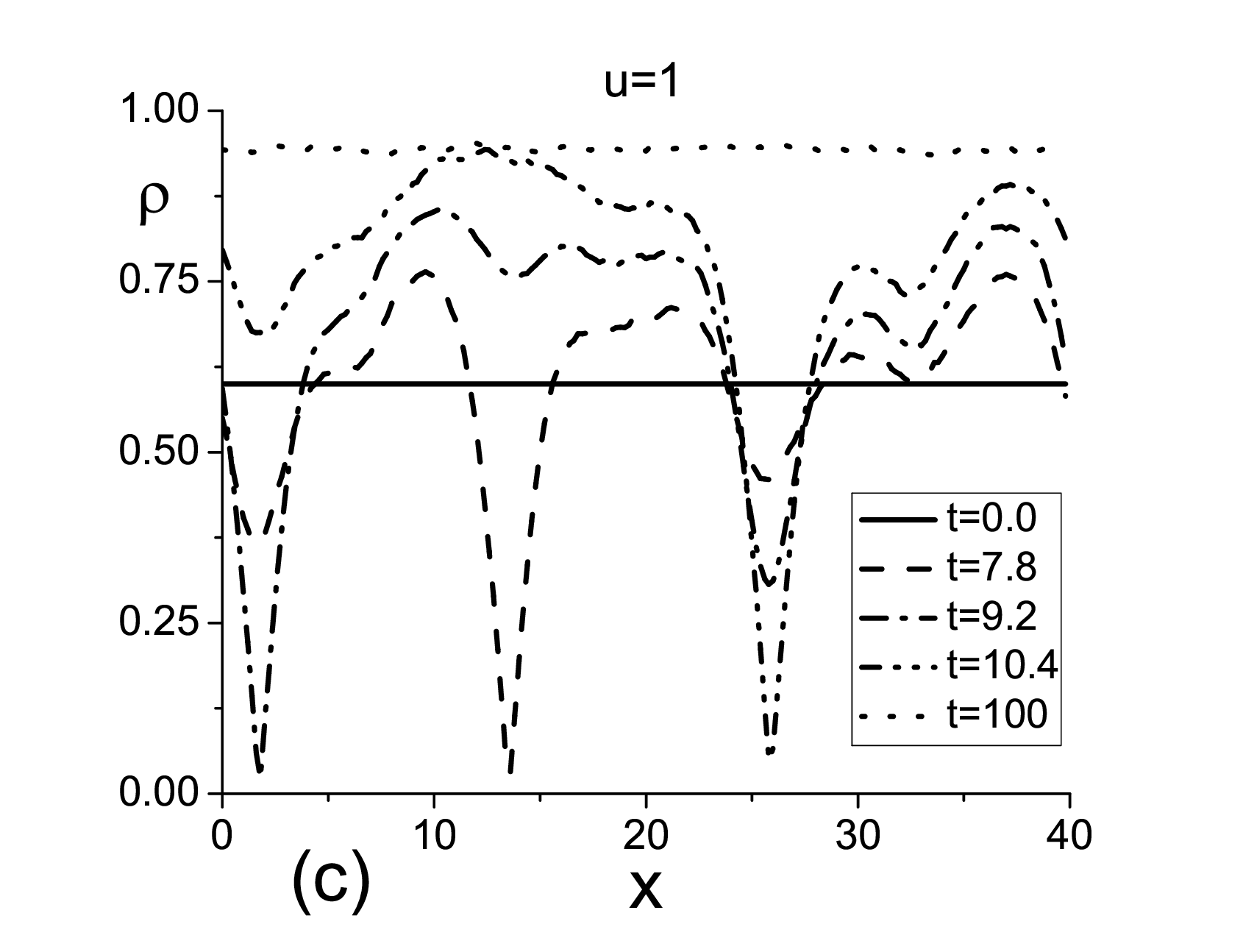}}
            {\includegraphics[width=7cm]{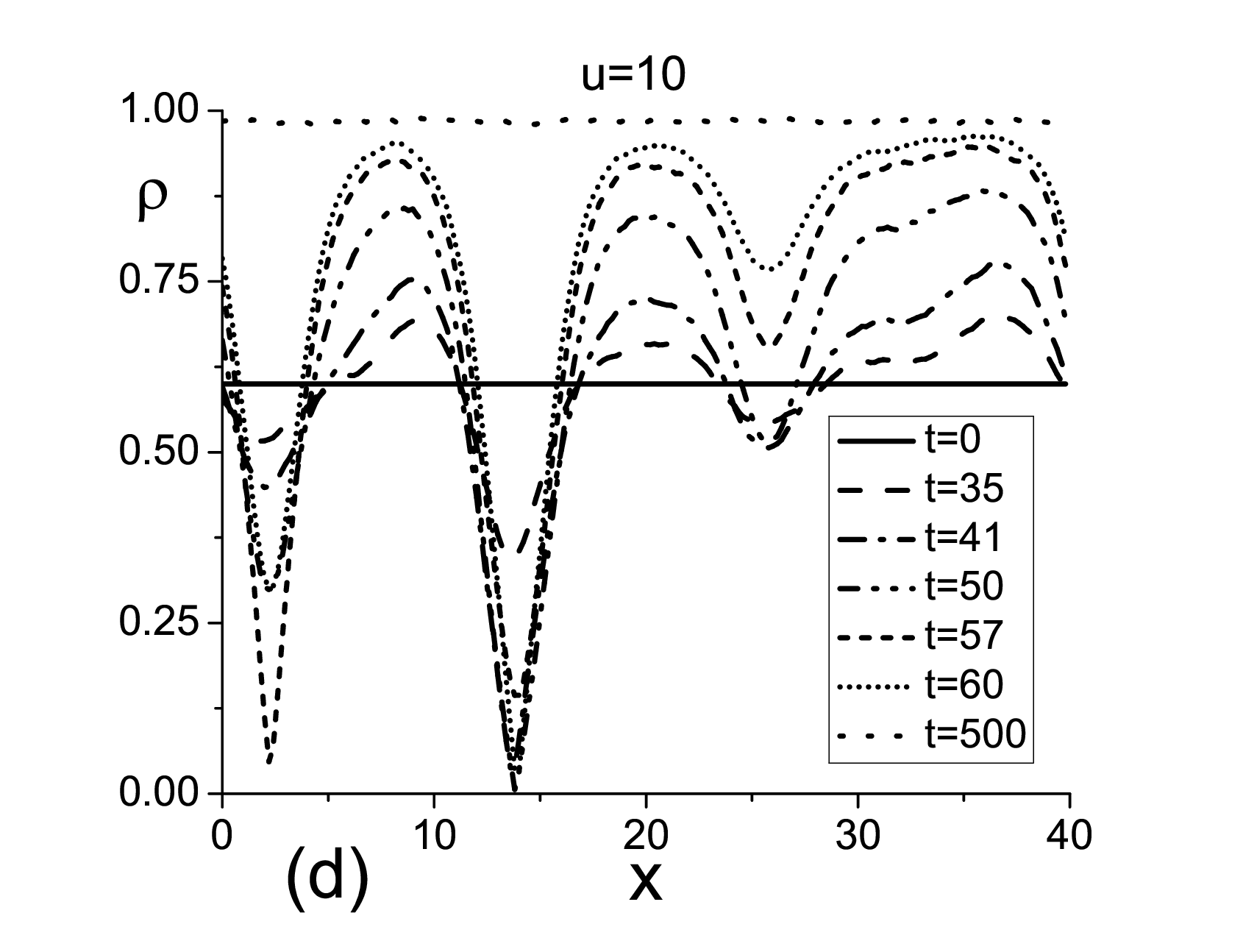}}
           \caption{Evolution of the distribution of the order parameter amplitude $\rho$ in the presence of a Langevin noise of the order of $10^{-1}$. The phase slip are accelerated and happen less simultaneously. For $u=10$ (d), we observe a mixing between consecutive and simultaneous phase slips. }
            \label{fig:multipleunoise}

\end{figure*}

As described in Eq. (\ref{eq:dtdgl1noise}), we add a Langevin noise. The amplitude of the noise corresponding to typical experimental values is on the order $10^{-3}$  which does not create any noticeable difference with the noiseless situation (all simulations excepting those shown in Fig. \ref{fig:multipleunoise} were performed using a Langevin noise of this amplitude). However, according to our simulations the increase in the level of the noise leads to a stronger instability. We notice that the phase slips are accelerated, especially in the beginning of the process. In the multiple phase slips case, the behavior of the order parameter can be significantly altered. The phase slips tend to happen ``less simultaneously'' and we even observe a mixing of consecutive and simultaneous phase slips [see Fig. \ref{fig:multipleunoise}(d)]. Rather surprisingly, the increase in the noise does not necessarily lead to the relaxation of the order parameter closer to the ground-state value. Sometimes we observe the opposite. For $u=0.5$, increasing the noise reduced the number of phase slips from four to three as one can see by comparing Figs. \ref{fig:multipleu}(b) and \ref{fig:multipleunoise}(b). We also note that the general conclusions of the stability analysis and the role of the parameter $u$ stay valid. Last but not the least, in the absence of noise, a nonuniform perturbation needs to be applied in order to initiate the phase slip process.

As pointed out in Ref. \onlinecite{VodoPeet02}, a strong magnetic field might trigger strong normal currents and destroy superconductivity because of heating effects. In our simulations, a strong magnetic field is indeed required to trigger simultaneous phase slips. However, in the $u<<1$ case, we observe simultaneous multiple phase slips already at magnetic fields on the order of $1 G$ for an Al sample of length $L=40\xi$ and $\xi=100$nm. This magnetic field is smaller than the value described as the maximum field sustainable by such a ring in Ref. \onlinecite{VodoDubo03}.
                The estimate of the heating of the ring may be made using Ohm's law, as we consider that during a phase slip event, a part of the ring of length $\xi$ behaves like a normal conductor of conductivity $\sigma_n$ during a time $\tau_{\theta}$ and sustains a current $J_s$ (in cgs units). The energy dissipated per unit of volume is then:
                \begin{equation}
                E=\frac{J_s^2}{\sigma_n}\tau_{\theta}=\frac{\left [\phi_0 A (1-A^2) \right ]^2}{16\pi^3 \xi^2\lambda_{\text{eff}}^2}.
                \end{equation}
                This energy is less than a quarter of the condensation energy $\frac{H_{c_2}^2}{16\pi\kappa^2}$ and such a small heating should not have any significant effect on the experimental detection of the phase slip. The case of multiple phase slips is more complicated and depends on the quality of the contacts of the sample with the thermostat. We believe that for the case of a free standing ring with a reasonable number of consecutive phase slips, the local temperature does not reach $T_c$ because of the large time scales involved [see Figs. \ref{fig:multipleu}(d) and \ref{fig:multipleu}(e)].
                 In the case of simultaneous phase slips, simulations confirm that the phase slip centers spread along the whole sample and therefore the distances are large in comparison with the heat diffusion length [see Figs. \ref{fig:multipleu}(a) and \ref{fig:multipleu}(b)]. In that case, the local heating is similar to the single phase slip case.

        \section{Conclusion}

        We formulated the stability condition for the superconducting state of a 1D ring in a constant magnetic field. Using the TDGL equations, we found out that the state is stable when the difference between the vorticity and the number of flux quanta enclosed in the ring is small. The relaxation towards the stable state must therefore involve one or more phase slips. The study of the multiple phase slips case revealed the importance of the ratio $u$ between the characteristic relaxation times of the amplitude and the phase of the order parameter. While $u<<1$ is favoring simultaneous phase slips, evolutions with $u>>1$ more often happen with consecutive phase slips. Therefore, $u>>1$ is often a necessary condition for the relaxation to the ground state. The effect of the Langevin noise present in the equations is also studied. It appears that for higher values of the noise, the behavior may be different; but the general conclusions of the study are still valid. In some cases, phase slips may be favored by local inhomogeneities which could favor the simultaneous phase slips case, but this goes beyond the scope of the present study.

\end{document}